\documentclass[12pt]{article}
\usepackage{graphicx}

      \usepackage[cp1251]{inputenc} 

\begin{document}
 \noindent {\footnotesize\it Astronomy Letters, 2013, Vol. 39, No. 11, pp. 753--758.}
 \newcommand{\dif}{\textrm{d}}

 \noindent
 \begin{tabular}{llllllllllllllllllllllllllllllllllllllllllllll}
 & & & & & & & & & & & & & & & & & & & & & & & & & & & & & & & & & & & & & \\\hline\hline
 \end{tabular}

 \vskip 1.0cm

 \centerline{\bf Orientation Parameters of the Cepheid System in the Galaxy}
 \bigskip
 \centerline{\bf V.V. Bobylev}
 \bigskip
 \centerline{\small \it Pulkovo Astronomical Observatory, St. Petersburg,  Russia}
 \centerline{\small \it Sobolev Astronomical Institute, St. Petersburg State University, Russia}
 \bigskip
 \bigskip
{\bf Abstract}—Based on the distribution of long-period Cepheids,
we have redetermined the orientation parameters of their principal
plane in the Galaxy. Based on 299 Cepheids with heliocentric
distances $r<20$~kpc and pulsation periods $P\geq5^d$, we have
found the directions of the three principal axes of the position
ellipsoid:
 $L_1=281.0\pm0.1^\circ,$ $B_1=-1.9\pm0.1^\circ,$
 $L_2= 11.0\pm0.7^\circ,$ $B_2=0.2\pm0.1^\circ$ and
 $L_3=275.9\pm0.7^\circ,$ $B_3=88.1\pm0.1^\circ$.
Thus, the line of nodes $l_\Omega=L_3+90^\circ=5.9^\circ$ is very
close to the direction to the Galactic center; the Cepheid
symmetry plane is inclined to the Galactic plane approximately by
$-2^\circ$ in the direction of the first axis ($L_1$). The
direction of the line of nodes found from old Cepheids ($P<5^d$)
differs significantly and is $l_\Omega=298^\circ$. The elevation
of the Sun above the Galactic plane has been estimated from 365
closer stars ($r<4$~kpc) without any constraint on the pulsation
period to be $h_\odot=23\pm5$~pc.


\section*{INTRODUCTION}
Cepheids play a very important role in studying the Galactic
structure. Their number increases; the calibration of the
period–luminosity relation needed to determine their distances is
improved. This has become possible owing to the fact that the
trigonometric parallaxes have been measured for several Cepheids
(Fouqu\'e et al. 2007). Using infrared photometry allowed the
interstellar extinction to be taken into account much more
accurately (Berdnikov et al. 2000).

The layer of neutral hydrogen in the Galaxy is known to be warped
at large Galactocentric distances (Westerhout 1957; Barton et al.
1988). Hydrogen rises above the Galactic plane in the second
quadrant and goes below it in the third and fourths quadrants. The
results of a study of this structure based on currently available
data on the HI and HII distributions are presented in Kalberla and
Dedes (2008) and Cersosimo et al. (2009), respectively. This
structure is revealed from the spatial distribution of stars and
dust (Drimmel and Spergel 2001), from the distribution of pulsars
(Yusifov 2004), from OB stars (Miyamoto and Zhu 1998), and from
the 2MASS red giant clump (Momany et al. 2006). The Cepheid system
exhibits a similar feature (Fernie 1968; Berdnikov 1987).

Several models were proposed to explain the nature of the Galactic
warp: (1) the interaction between the disk and a nonspherical dark
matter halo (Sparke and Casertano 1988); (2) the gravitational
influence from the closest satellites of the Galaxy (Bailin 2003);
(3) the interaction of the disk with the circumgalactic flow of
high-velocity hydrogen clouds produced by mass transfer between
the Galaxy and the Magellanic Clouds (Olano 2004); (4) the
intergalactic flow (L\'opez-Corredoira et al. 2002); and (5) the
interaction with the intergalactic magnetic field (Battaner et al.
1990).

The goal of this paper is to redetermine the orientation of the
Cepheid system in our Galaxy. For this purpose, we apply a method
that allows this problem to be solved in a general form. Studying
the dependence of the derived parameters on the stellar age and
distance is also of great importance.

\section*{DATA}
Here, we use Cepheids of the Galaxy’s flat component classified as
DCEP, DCEPS, CEP(B), CEP in the GCVS (Kazarovets et al. 2009) as
well as CEPS used by other authors. To determine the distance
based on from the period–luminosity relation, we used the
calibration from Fouqu\'e et al. (2007):
 $\langle M_V\rangle=-1.275-2.678 \log P,$ where the period $P$ is in days.
Given $\langle M_V\rangle,$ taking the period-averaged apparent
magnitudes $\langle V\rangle$ and extinction
 $A_V=3.23 E(\langle B\rangle-\langle V\rangle)$ mainly from Acharova et al. (2012)
and, for several stars, from Feast and Whitelock (1997), we
determine the distance $r$ from the relation
 \begin{equation}\displaystyle
 r=10^{\displaystyle -0.2(\langle M_V\rangle-\langle V\rangle-5+A_V)}.
 \label{Ceph-02}
 \end{equation}
For a number of Cepheids (without extinction data), we used the
distances from the catalog by Berdnikov et al. (2000), which were
determined from infrared photometry.

With the goals of our study in mind, we concluded that it was
better not to use several stars lying higher than 2 kpc above the
Galactic plane and those located deep in the Galaxy's inner
region. Thus, we used two limitations,
 \begin{equation}
    |Z|<2~\hbox {kpc}, \quad   X<6~\hbox {kpc},
 \label{criterii-xz}
 \end{equation}
satisfied by 465 Cepheids. Their distributions in projections onto
the Galactic $XY,$ $XZ,$ and $YZ$ planes are shown in Figs. 1--3.

Several calibrations proposed to estimate the Cepheid ages are
known. Here,we use the calibration by Efremov (2003),
 \begin{equation}
 \log t=8.50-0.65 \log P,
 \label{AGE-EFREM}
 \end{equation}
derived from Cepheids belonging to open clusters of the Large
Magellanic Cloud.

\section*{THE METHOD}
We apply the well-known method of determining the symmetry plane
of a stellar system with respect to the principal (in our case,
Galactic) coordinate system. The basics of this approach were
described by Polak (1935), and the technique for estimating the
errors in the angles can be found in Parenago (1951) and
Pavlovskaya (1971).

In the rectangular coordinate system centered on the Sun, the $x$
axis is directed toward the Galactic center, the $y$ axis is in
the direction of Galactic rotation $(l=90^\circ, b=0^\circ),$ and
the $z$ axis is directed toward the North Galactic Pole. Then,
 \begin{equation}
  \begin{array}{rll}
  x&=&r\cos l\cos b,\\
  y&=&r\sin l\cos b,\\
  z&=&r\sin b.
   \label{ff-1}
  \end{array}
 \end{equation}
Let $m, n,$ and $k$ be the direction cosines of the pole of the
sought-for great circle from the $x, y,$ and $z$ axes. The
sought-for symmetry plane of the stellar system is then determined
as the plane for which the sum of the squares of the heights,
$h=mx+ny+kz,$ is at a minimum:
 \begin{equation}
 \sum h^2=\hbox {min}.
 \label{ff-2}
 \end{equation}
The sum of the squares
 \begin{equation}
 h^2=x^2m^2+y^2n^2+z^2k^2+2yznk+2xzkm+2xymn
 \label{ff-3}
 \end{equation}
can be designated as $2P=\sum h^2.$ As a result, the problem is
reduced to searching for the minimum of the function $P:$
 \begin{equation}
 2P=am^2+bn^2+ck^2+2fnk+2ekm+2dmn,
 \label{ff-4}
 \end{equation}
where the second-order moments of the coordinates
 $a=[xx],$
 $b=[yy],$
 $c=[zz],$
 $f=[yz],$
 $e=[xz],$
 $d=[xy],$  written via the
Gauss brackets, are the components of a symmetric tensor:
 \begin{equation}
 \left(\matrix {
  a& d & e\cr
  d& b & f\cr
  e& f & c\cr }\right),
 \label{ff-5}
 \end{equation}
whose eigenvalues $\lambda_{1,2,3}$ are found from the solution of
the secular equation
 \begin{equation}
 \left|\matrix
 {
a-\lambda&          d&        e\cr
       d & b-\lambda &        f\cr
       e &          f&c-\lambda\cr
 }
 \right|=0,
 \label{ff-7}
 \end{equation}
and the directions of the principal axes $L_{1,2,3}$ and
$B_{1,2,3}$ are found from the relations
 \begin{equation}
 \renewcommand{\arraystretch}{2.2}
  \begin{array}{lll}
  \displaystyle
 \tan L_{1,2,3}={{ef-(c-\lambda)d}\over {(b-\lambda)(c-\lambda)-f^2}},\\
 \displaystyle
 \tan B_{1,2,3}={{(b-\lambda)e-df}\over{f^2-(b-\lambda)(c-\lambda)}}\cos L_{1,2,3}.
 \label{ff-42}
  \end{array}
 \end{equation}
The errors in $L_{1,2,3}$ and $B_{1,2,3}$ are estimated according
to the following scheme:
 \begin{equation}
 \renewcommand{\arraystretch}{2.2}
  \begin{array}{lll}
  \displaystyle
 \varepsilon (L_2)= \varepsilon (L_3)= {{\varepsilon (\overline {xy})}\over{a-b}},\\
  \displaystyle
 \varepsilon (B_2)= \varepsilon (\varphi)={{\varepsilon (\overline {xz})}\over{a-c}},\\
  \displaystyle
 \varepsilon (B_3)= \varepsilon (\psi)= {{\varepsilon (\overline {yz})}\over{b-c}},\\
  \displaystyle
 \varepsilon^2 (L_1)={\varphi^2 \varepsilon^2 (\psi)+\psi^2 \varepsilon^2 (\varphi)\over{(\varphi^2+\psi^2)^2}},\\
  \displaystyle
 \varepsilon^2 (B_1)= {\sin^2 L_1 \varepsilon^2 (\psi)+\cos^2 L_1 \varepsilon^2 (L_1)\over{(\sin^2 L_1+\psi^2)^2}},
 \label{ff-65}
  \end{array}
 \end{equation}
where
 $$
 \varphi=\cot B_1 \cos L_1, \quad \psi=\cot B_1 \sin L_1.
 $$
The three quantities $\overline {x^2y^2}$, $\overline {x^2z^2}$
and $\overline {y^2z^2},$ should be calculated in advance. Then,
 \begin{equation}
 \renewcommand{\arraystretch}{1.6}
  \begin{array}{lll}
  \displaystyle
 \varepsilon^2 (\overline {xy})= (\overline{x^2y^2}-d^2)/n, \\
  \displaystyle
 \varepsilon^2 (\overline {xz})= (\overline {x^2z^2}-e^2)/n, \\
  \displaystyle
 \varepsilon^2 (\overline {yz})= (\overline {y^2z^2}-f^2)/n,
 \label{ff-73}
  \end{array}
 \end{equation}
where $n$ is the number of stars. Thus, the algorithm for solving
the problem consists in setting up the function $2P$~(7), seeking
for the roots of the secular equation (9), whose specific values
are of no interest to us, and estimating the directions of the
principal axes of the position ellipsoid from Eqs. (10)--(12). In
the classical case, the problem was solved for a unit sphere
($r=1$), but here we propose to use the distances (which act as
the weights).

\begin{figure}[t]
{\begin{center}
 \includegraphics[width=0.9\textwidth]{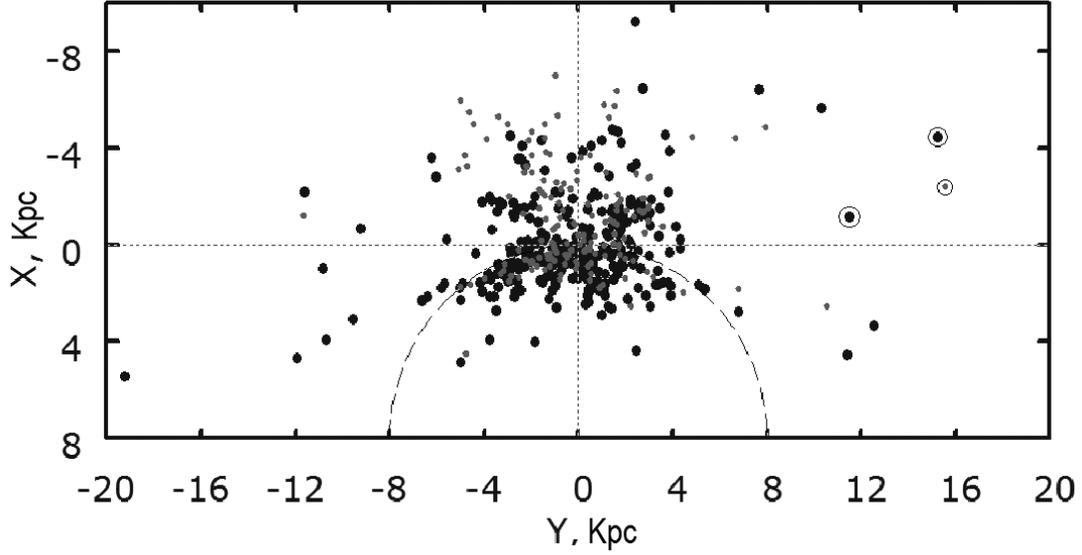}
 \caption{
Distribution of Cepheids in projection onto the Galactic $XY$
plane. The Sun is at the intersection of the dotted lines; the
dashed line indicates the circle of radius $R_0=8$~kpc around the
Galactic center. The filled circles are long-period Cepheids
$(P\geq5^d);$ the small gray circles are short-period Cepheids
$(P<5^d);$ the circles mark the three Cepheids with large $Z$
discussed in the text. }
 \label{f1}
\end{center}}
\end{figure}
\begin{figure}[t]
{\begin{center}
 \includegraphics[width=0.7\textwidth]{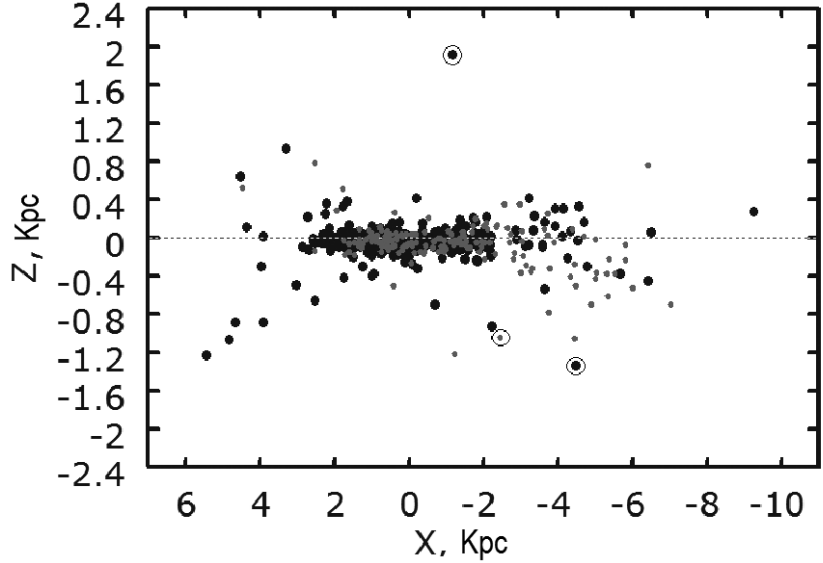}
 \caption{
Distribution of Cepheids in projection onto the Galactic $XZ$
plane. The Galactic center is on the left (at $X=8$~kpc). The
notation is the same as that in Fig.~1.}
 \label{f2}
\end{center}}
\end{figure}
\begin{figure}[t]
{\begin{center}
 \includegraphics[width=0.7\textwidth]{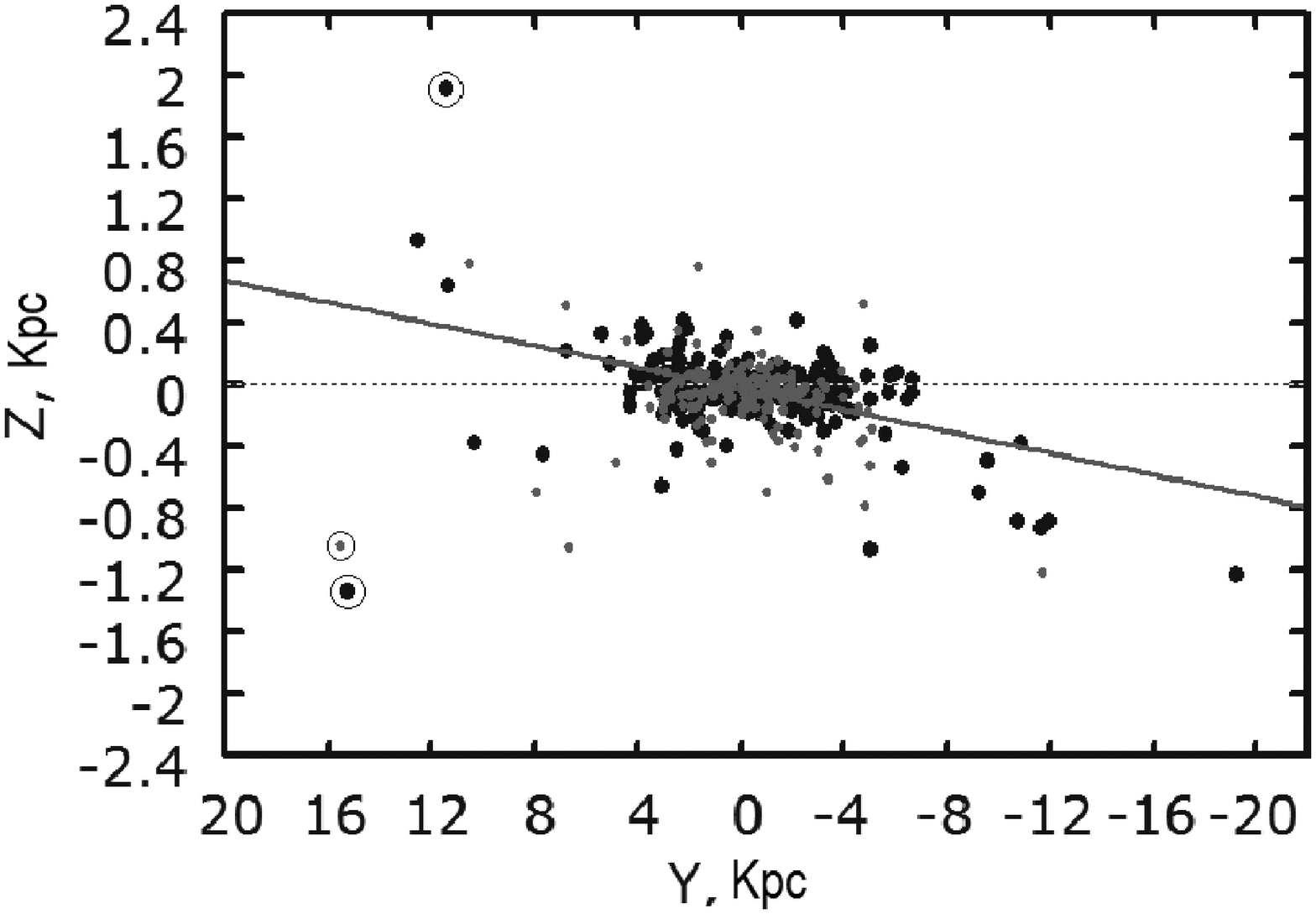}
 \caption{
Distribution of Cepheids in projection onto the Galactic $YZ$
plane. The slope of the solid line is $2^\circ$. The notation is
the same as that in Fig.~1.}
 \label{f3}
\end{center}}
\end{figure}

\section*{RESULTS}
Based on the entire sample of Cepheids (465 stars), whose
distribution in the Galaxy is shown in Figs. 1--3, we found the
following directions of the principal axes of the position
ellipsoid:
 \begin{equation}
 \renewcommand{\arraystretch}{1.2}
  \matrix {
  L_1=278.96\pm0.05^\circ, & B_1=-1.33\pm0.00^\circ, \cr
  L_2=~~8.93\pm0.43^\circ, & B_2=~1.41\pm0.04^\circ, \cr
  L_3=232.37\pm0.43^\circ, & B_3=88.06\pm0.07^\circ. \cr
  }
 \label{rezult-1}
 \end{equation}
The mean age for this sample of Cepheids is
 ${\overline t}=98$~Myr. According to solution~(13), the slope of the solid
line in Fig.~3 corresponds to $(90^\circ-B_3).$ For comparison, we
present the results of our calculations for the distribution on a
unit sphere $(r=1)$ obtained using the same stars:
 \begin{equation}
  \matrix {
  L_1=283.32\pm0.02^\circ, & B_1=-0.66\pm0.00^\circ, \cr
  L_2=~13.32\pm0.67^\circ, & B_2=-0.37\pm0.03^\circ, \cr
  L_3=312.95\pm0.67^\circ, & B_3=89.24\pm0.03^\circ, \cr
  }
 \label{rezult-2}
 \end{equation}
which differ significantly from the parameters of solution (13).
Although the errors in the unknowns $B_{1,2,3}$ in solution (14)
are smaller, the angles $B_{1,2,3}$ are also considerably smaller
than those in solution (13). However, it can be clearly seen from
Fig.~3 that the slope of $2^\circ$ is better applicable to the
data than $\approx$0.5$^\circ$ (as follows from solution~(14)).
This discrepancy decreases if the stars are considered in narrow
distance ranges. Below, we consider only the results of the
solutions with distances.

Distant stars make a major contribution to solution~(13) (they
have the largest weights). The solutions obtained from distant
stars ($3<r<20$~kpc) with different pulsation periods (ages) are
of interest. For the youngest stars,
 \begin{equation}
  \matrix {
  L_1=279.6\pm0.3^\circ, & B_1=-2.1\pm0.1^\circ, \cr
  L_2=~~9.9\pm1.3^\circ, & B_2=~0.6\pm0.1^\circ, \cr
  L_3=262.8\pm1.3^\circ, & B_3=87.8\pm0.3^\circ, \cr
  &P\geq9^d, \cr
  &n_\star=63, \cr
  &{\overline t}=54~\hbox {Myr}; \cr
  }
 \label{rezult-3}
 \end{equation}
for middle-age stars,
 \begin{equation}
  \matrix {
  L_1=284.5\pm0.2^\circ, & B_1=-1.4\pm0.1^\circ, \cr
  L_2=~14.5\pm2.2^\circ, & B_2=~0.3\pm0.1^\circ, \cr
  L_3=272.3\pm2.2^\circ, & B_3=88.5\pm0.2^\circ, \cr
  &5^d\leq P<9^d, \cr
  &n_\star=51, \cr
  &{\overline t}=96~\hbox {Myr}; \cr
  }
 \label{rezult-4}
 \end{equation}
for the oldest stars:
 \begin{equation}
  \matrix {
  L_1=261.6\pm0.4^\circ, & B_1=-0.9\pm0.0^\circ, \cr
  L_2=351.6\pm1.0^\circ, & B_2=~3.2\pm0.4^\circ, \cr
  L_3=188.0\pm1.0^\circ, & B_3=86.7\pm0.1^\circ, \cr
  &P<5^d, \cr
  &n_\star=63, \cr
  &{\overline t}=133~\hbox {Myr}. \cr
  }
 \label{rezult-5}
 \end{equation}
In contrast to the results of solutions (15)--(16), which, on the
whole, agree between themselves, the oldest Cepheids give a
significantly different orientation of the line of nodes,
$l_\Omega=L_3+90^\circ=278^\circ.$ This is no surprise, because
old Cepheids have had time to recede from their birthplace, they
made more than half of their revolution around the Galactic
center, i.e., they were formed in a different part of the Galaxy
(for example, with respect to the Magellanic Clouds). The
surprising thing is that a slope of $\approx3^\circ$ is present in
their distribution (the angles $B_2$ and $B_3$). This finding may
imply that the disk warp can be a long-lived structure, at least
older than $\approx150$~Myr.

The roots of the secular equation~(9) describe the shape of the
ellipsoid but provide no information about the coordinates of its
center. The shift along the $z$ coordinate, which characterizes
the elevation of the Sun above the Galactic plane
$h_\odot=-{\overline z},$ is most interesting.

Three stars with very large $Z$ are marked in Figs. 1--3. These
are two long-period Cepheids,
 DR~Cep ($P =19.8^d, Z = 1.9$~kpc) and
 FQ~Lac ($P =11.3^d, Z =-1.3$~kpc), and one short-period Cepheid,
 IT~Lac ($P = 2.6^d, Z =-1.0$~kpc). Unfortunately, as yet no information
about their radial velocity measurements is available for them.
Their proper motions from the UCAC4 catalog (Zacharias et al.
2013) are very unreliable. This is because at such large distances
($r\approx15$~kpc), a typical error $e_\mu\approx2$~mas yr$^{-1}$
will contribute $4.74 r e_\mu\approx140$~km s$^{-1}$ to the space
velocity, which is very much. Therefore, it is not yet possible to
judge the character of their velocities. These stars most likely
have peculiar velocities. For this reason, we decided not to use
these stars to determine the orientation parameters.

Since the results of solutions (15)--(16) do not differ greatly,
the interval of periods can be combined. Based on 299 stars
 $(r<20$~kpc) with pulsation periods $P\geq5^d$ (${\overline t}=77$~Myr), we have now
found
 \begin{equation}
  \matrix {
  L_1=281.0\pm0.1^\circ, & B_1=-1.9\pm0.1^\circ, \cr
  L_2=~11.0\pm0.7^\circ, & B_2=~0.2\pm0.1^\circ, \cr
  L_3=275.9\pm0.7^\circ, & B_3=88.1\pm0.2^\circ, \cr
  }
 \label{rezult-66}
 \end{equation}
the line of nodes $l_\Omega=L_3+90^\circ=5.9^\circ$ is close to
the direction to the Galactic center. The parameters~(18) agree
satisfactorily with the results of analyzing the layer of neutral
hydrogen (Kalberla and Dedes 2008). It should be noted that the
distances to hydrogen clouds are estimated from the radial
velocities (kinematic distances) with a low accuracy; in our case,
however, the accuracy is higher, because the distance error is, on
average, 10--15\%. Therefore, our results are of indubitable
interest.

Based on 163 stars ($r<20$~kpc) with pulsation periods $P<5^d$
(${\overline t}=138$~Myr), we found
 \begin{equation}
  \matrix {
  L_1=249.5\pm0.4^\circ, & B_1=-2.1\pm0.1^\circ, \cr
  L_2=339.4\pm1.9^\circ, & B_2=~1.9\pm0.2^\circ, \cr
  L_3=208.1\pm1.9^\circ, & B_3=87.2\pm0.1^\circ, \cr
  }
 \label{rezult-77}
 \end{equation}
the direction of the line of nodes is $l_\Omega=298^\circ$.

The elevation of the Sun above the Galactic plane $h_\odot$
depends on the heliocentric distance, which is clearly seen from
Figs. 2, 3. Our calculations show that a sample with a radius of
4--5~kpc is optimal (the error in $h_\odot$ is smallest). For
example, based on a sample of the closest (71 stars) Cepheids from
the range $r\leq1$~kpc (with the rejection according to the
$3\sigma$ criterion), we found
 \begin{equation}
 h_\odot=30\pm9~\hbox {pc},
 \label{rezult-7}
 \end{equation}
but the influence of nonuniformities in the distribution of stars
is great here. Based on 365 stars from the range $r\leq4$~kpc, we
found
 \begin{equation}
 h_\odot=23\pm5~\hbox {pc},
 \label{rezult-8}
 \end{equation}
while based on the remaining 100 stars from the range
4~kpc$<r<20$~kpc, we found
 \begin{equation}
 h_\odot=45\pm39~\hbox {pc}.
 \label{rezult-9}
 \end{equation}

\section*{DISCUSSION}
Fernie (1968) estimated the direction of the line of nodes for the
Cepheid system, $7^\circ\pm4^\circ,$ the inclination
$-0.8^\circ\pm0.2^\circ$ in the direction $l=277^\circ,$ and the
elevation of the Sun above the Galactic plane,
$h_\odot=45\pm15$~pc from 328 stars. Since the sample of stars in
Fernie (1968) probably contained quite a few old Cepheids, the
inclination turned out to be small, and the determination of
$h_\odot$ was affected by distant Cepheids. Berdnikov (1987) found
$h_\odot=26\pm6$~pc from 363 stars, in good agreement with our
result (21). Our value of $h_\odot$ (21) is also in good agreement
with $h_\odot=17\pm3$~pc obtained by Joshi (2007) from young open
star clusters and OB stars.

A discussion of the results of analyzing the warp of the hydrogen
layer obtained from neutral, ionized, and molecular hydrogen can
be found, for example, in Cersosimo et al. (2009). Hydrogen
reaches its maximum elevations above the Galactic plane z = from
$+$300 to $+$400 pc (at $R\approx$12~kpc) in the first and second
quadrants and $z=$ from $-150$ to $-200$~pc in the third and
fourth quadrants (i.e., the warp is nonlinear). It can be seen
from Fig.~3 that even after the elimination of the above three
stars, the dispersion of the positions is larger at positive $y$
(on the left in the Fig.~3); on average, the heights of the stars
are larger than those of hydrogen. At positive $y,$ there are also
Cepheids with negative $z.$ This can be related to their ages;
having been formed $\approx$50~Myr ago, they could be displaced
below the plane in such a time. To confirm this assumption, it is
necessary to analyze the space velocities of Cepheids, which we
plan to do in another paper.

On the whole, we can conclude that the connection of Cepheids with
the Galactic warp is beyond doubt.

\section*{CONCLUSIONS}
Based on the distribution of Cepheids, we redetermined the
orientation parameters of their principal plane in the Galaxy.
Based on 299 stars at heliocentric distances $r<20$~kpc with
pulsation periods $P\geq5^d$ we found the directions of the three
principal axes of the position ellipsoid (solution (18)):
 $L_1=281.0\pm0.1^\circ,$ $B_1=-1.9\pm0.1^\circ,$
 $L_2= 11.0\pm0.7^\circ,$ $B_2=0.2\pm0.1^\circ$ and
 $L_3=275.9\pm0.7^\circ,$ $B_3=88.1\pm0.1^\circ.$ The line of
nodes $l_\Omega=L_3+90^\circ=5.9^\circ$ is very close to the
direction to the Galactic center; the Cepheid symmetry plane is
inclined to the Galactic plane approximately by $-2^\circ$ in the
direction of the first axis $(L_1).$

The oldest Cepheids (163 stars at $r<20$~kpc with pulsation
periods $P<5^d)$ give a significantly different orientation of the
line of nodes (solution (19)):
 $L_1=249.5\pm0.4^\circ, B_1=-2.1\pm0.1^\circ,$
 $L_2=339.4\pm1.9^\circ, B_2=~1.9\pm0.2^\circ$ and
 $L_3=208.1\pm1.9^\circ, B_3=87.2\pm0.1^\circ,$
The direction of the line of nodes $l_\Omega=298^\circ$ differs
approximately by $65^\circ$ from that obtained from a sample of
younger Cepheids.

The Sun's elevation above the Galactic plane was estimated from
365 stars at $r<4$~kpc without any constraint on the pulsation
period $P$ to be $h_\odot=23\pm5$~pc.

In future, the method considered here can be useful for analyzing
large amounts of data, for example, those from the GAIA space
experiment or masers with their trigonometric parallaxes measured
by means of VLBI.

\subsection*{ACKNOWLEDGMENTS} We are grateful to the referee for
helpful remarks that contributed to a improvement of the paper.
This work was supported by the ``Nonstationary Phenomena in
Objects of the Universe'' Program of the Presidium of the Russian
Academy of Sciences and the ``Multiwavelength Astrophysical
Research'' grant no. NSh--16245.2012.2 from the President of the
Russian Federation.

\section*{REFERENCES}
{\small

\quad~~1. A.A. Acharova, Yu.N. Mishurov, and V.V. Kovtyukh, Mon.
Not. R. Astron. Soc. 420, 1590 (2012).

2. J. Bailin, Astrophys. J. 583, L79 (2003).

3. E. Battaner, E. Florido, and M.L. Sanchez-Saavedra, Astron.
Astrophys. 236, 1 (1990).

4. L.N. Berdnikov, Astron. Lett. 13, 45 (1987).

5. L.N. Berdnikov, A.K. Dambis, and O.V. Vozyakova, Astron.
Astrophys. Suppl. Ser. 143, 211 (2000).

6. W.B. Burton, in Galactic and Extragalactic Radio Astronomy, Ed.
by G. Verschuur and K. Kellerman (Springer, New York, 1988), p.
295.

7. J.C. Cersosimo, S. Mader, N.S. Figueroa, et al., Astrophys. J.
699, 469 (2009).

8. R. Drimmel and D.N. Spergel, Astrophys. J. 556, 181 (2001).

9. Yu.N. Efremov, Astron.Rep. 47, 1000 (2003).

10. M. Feast and P. Whitelock, Mon. Not. R. Astron. Soc. 291, 683
(1997).

11. J.D. Fernie, Astron. J. 73, 995 (1968).

12. P. Fouqu, P. Arriagada, J. Storm, et al., Astron. Astrophys.
476, 73 (2007).

13. Y.C. Joshi, Mon. Not. R. Astron. Soc. 378, 768 (2007).

14. P.M.W. Kalberla and L. Dedes, Astron. Astrophys. 487, 951
(2008).

15. E.V. Kazarovets, N.N. Samus, O.V. Durlevich, et al., Astron.
Rep. 53, 1013 (2009).

16. M. L\'opez-Corredoira, J. Betancort-Rijo, and J. Beckman,
Astron. Astrophys. 386, 169 (2002).

17. M. Miyamoto and Z. Zhu, Astron. J. 115, 1483 (1998).

18. Y. Momany, S. Zaggia, G. Gilmore, et al., Astron. Astrophys.
451, 515 (2006).

19. C.A. Olano, Astron. Astrophys. 423, 895 (2004).

20. P.P. Parenago, Trudy Gos. Astron. Inst. Shternberga 20, 26
(1951).

21. E.D. Pavlovskaya, in Practical Works in Stellar Astronomy, Ed.
by P.G. Kulikovskii (Nauka, Moscow, 1971), p. 162 [in Russian].

22. I.F. Polak, Introduction to Stellar Astronomy (ONTI, Moscow,
Leningrad, 1935) [in Russian].

23. L. Sparke and S. Casertano, Mon. Not. R. Astron. Soc. 234, 873
(1988).

24. G. Westerhout, Bull. Astron. Inst. Netherlands 13, 201 (1957).

25. I. Yusifov, astro-ph: 0405517 (2004).

26. N. Zacharias, C. Finch, T. Girard, et al., Astron. J. 145, 44
(2013).
}

\end{document}